\documentstyle[12pt,aaspp4]{article}
\newcommand{\cii}{\ion{C}{2}}
\newcommand{\civ}{\ion{C}{4}}
\newcommand{\oi}{\ion{O}{1}}

\newcommand{\ea}{{et~al.}}
\newcommand{\hi}{\ion{H}{1}}
\newcommand{\IUE}{{\it IUE}}
\newcommand{\kms}{km s$^{-1}$}
\newcommand{\lbol}{ {\em l$_{bol}$}}
\newcommand{\logg}{$\log g$}
\newcommand{\lya}{\mbox{Ly$\alpha$}}

\newcommand{\msun}{M$_{\sun}$}
\newcommand{\mv}{{M}$_{V}$ }

\newcommand{\teff}{T$_{eff}$}
\newcommand{\vsini}{$v \sin i$}

\begin{document}

\title{The White Dwarf Companions of 56 Persei and HR 3643}

\author{Wayne Landsman\altaffilmark{1}}
\affil{Hughes STX, NASA/GSFC, Code 681,
Greenbelt, MD 20771}
\affil{Electronic mail: landsman@stars.gsfc.nasa.gov}

\author{Theodore Simon\altaffilmark{1}}
\affil{Institute for Astronomy, University of Hawaii,
2680 Woodlawn Drive, Honolulu, HI 96822}
\affil{Electronic mail: simon@uhifa.ifa.hawaii.edu}

\author{P. Bergeron}
\affil{D\'{e}partement de Physique, Universit\'{e} de Montr\'{e}al,
C.P. 6128, Succ. Centre-Ville,}
\affil{Montr\'{e}al, Qu\'{e}bec, Canada, H3C~3J7}
\affil{Electronic mail: bergeron@astro.umontreal.ca}

\altaffiltext{1}{Guest Observer with the {\em International Ultraviolet
Explorer (IUE)} satellite. }
\begin{center}
to be published in PASP (March 1996)
\end{center}
\vspace{0.1in}

\begin{abstract}

We have obtained low-dispersion \IUE\ spectra of the stars 56 Persei (F4~V)
and HR 3643 (F7~II), as part of a survey of late-type stars with a 1565~\AA\
flux excess in the {\it TD-1} ultraviolet sky survey.
The \IUE\ spectrum of each star reveals the presence of a hot white dwarf
companion.  We fit the \lya\ profile and ultraviolet continuum using
pure hydrogen models, but the distance of the primary star
is also needed to uniquely constrain the white dwarf parameters.
We derive \teff\ = $16,420 \pm 420$ K, \logg = $8.46 \pm 0.2$ for
the white dwarf companion to 56 Per, using the photometric distance of
$30.1 \pm 2.8$ pc.    The implied white dwarf mass is $0.90 \pm 0.12$ \msun,
considerably above the median mass ($\sim 0.6$ \msun) of single white dwarfs.
The parameters of the white dwarf in HR 3643 are not well constrained, mainly
due to a large uncertainty in the distance.
By assuming a reasonable range of gravity for the white dwarf
($7.3 <$ \logg $< 9.0$), we derive $-1.4 <$ \mv $< 0.6$ for the F7~II star,
and $28,970 <$ \teff $< 35,990$ K for the white dwarf.

Prompted by our detection of a white dwarf companion of a luminous F star, we
have examined the \IUE\ archives to assess the upper limits on possible
white dwarf companions to Cepheids.    The detection of a Cepheid -- white
dwarf binary would provide important insights concerning the most massive
progenitors of white dwarfs.
Only for the cases of $\alpha$ UMi and $\beta$ Dor are  existing \IUE\
spectra of Cepheids sufficiently deep to rule out the presence of a white dwarf
companion.

\end{abstract}

\section{Introduction}

Fewer than twenty stars in the Yale Bright Star Catalog are known
to have white dwarf companions.   This number is almost certainly
underestimates the true binary fraction, due to  the observational
difficulty of detecting white dwarf companions of bright (\mv $\lesssim 7$)
stars.    The white dwarf companion
can be spatially resolved in the classical nearby visual binaries,
such as Procyon and Sirius, as well as in common proper motion systems such as
HR 6094/CD $-38\arcdeg10980$ (\cite{oswalt88}).
In unresolved systems, the white dwarf is generally too faint to be revealed
by optical spectroscopy, but may be detected at shorter
wavelengths, provided that the white dwarf is hotter than the primary star.
Several white dwarf companions of late-type stars have been serendipitously
discovered using the {\em International Ultraviolet Observer (\IUE)} satellite
(e.g.\ \markcite{bohm92,bohm93}B\"ohm-Vitense 1992, 1993), but there
has been no systematic ultraviolet survey for such systems.
The determination of the fraction of non-degenerate stars with
white dwarf companions would have several important astrophysical applications.
As one example, the white dwarf binary fraction might provide a significant
correction to the single white dwarf luminosity function, which can be used
to estimate the star formation history of the Galaxy (\cite{wood92}).

An important breakthrough in the detection of white dwarf
binaries was provided by the ROSAT/WFC and EUVE all-sky surveys,
with
the discovery (thus far) of ten A -- K stars with white dwarf companions
(\cite{bars94,vennes95}).
But the EUV surveys suffer from three important selection effects that limit
their utility for the derivation of the white dwarf binary fraction.    First,
the EUV surveys are sensitive only to DA white dwarfs with
\teff $\gtrsim$ 24,000 K.
Second, the EUV flux is strongly attenuated by the local interstellar medium,
resulting in an asymmetric distribution of detected sources (\cite{warwick93}).
Third, the EUV flux can be strongly suppressed by the presence of
trace absorbers  (helium or metals) in the white dwarf photosphere.
The presence of trace absorbers is believed to be the
main reason why the number of white dwarfs discovered in the ROSAT survey is
only about one-tenth that predicted in pre-flight models (\cite{flem93}).

An all-sky {\em ultraviolet} ($\sim 1400$ \AA) survey of late-type stars
for white dwarf companions would suffer minimal selection effects due to
ultraviolet extinction or to the presence of trace photospheric absorbers,
and could reveal the presence of
DA and non-DA white dwarf companions as cool as $\sim 10,000$ K.
Unfortunately, sensitive
ultraviolet imaging experiments such as
FAUST (\cite{bow95a}) or the Ultraviolet Imaging Telescope (UIT,
\cite{stech92}) have covered only a small fraction of the sky.
In 1971, the S2/68 ultraviolet sky survey telescope on the {\it TD-1}
satellite did survey the entire sky at 1565 \AA\ (\cite{thompson78}), but only
to a limiting sensitivity of about 10$^{-12}$
erg cm$^{-2}$ s$^{-1}$ \AA$^{-1}$, and using a
large $11'$ by $18'$ entrance slot that presented severe problems with source
confusion.    Nevertheless, the utility of the {\it TD-1} catalog for the
detection of hidden white dwarf binaries was recently demonstrated for the case
of HR 1608 (K0 IV, V = 5.4).    The 1565 \AA\
flux for HR 1608 listed by \markcite{thompson78} Thompson \ea\ is
$1.1 \pm 0.1 \times 10^{-12}$ erg cm$^{-2}$ s$^{-1}$ \AA$^{-1}$, which is far
above what is expected for a K0 IV star.
However, HR 1608 was not observed with IUE
until after its detection as an EUV source
in the ROSAT WFC catalog (\cite{lsb93}).    The \IUE\
spectrum clearly shows the presence of a white dwarf companion, and verifies
the accuracy of the {\it TD-1} flux.

We are currently using \IUE\ to observe late-type stars with a UV
excess at 1565~\AA\ recorded in the
{\it TD-1} catalog.   In this paper, we report the detection of white dwarf
companions to the V = 5.8, F4 V star 56 Persei (=HR 1379, HD 27786) and the
V=4.5, F7 II star HR 3643 (= HD 78791).
The detection of a white dwarf companion to a luminous F star
is particularly interesting, since such a system implies a massive ($>$ 2
\msun)
progenitor for the white dwarf.   In the
case of HR 3643, however, we find that the constraints that can be placed on
the white dwarf progenitor are limited by the difficulty of determining the
distance, mass, and age of the non-variable bright giant primary.
These difficulties prompted us to consider the feasibility of detecting
white dwarf companions to Cepheids, where the distance, mass, and age can
be derived to much better precision.   Therefore, in section 4, we report on
a search of the \IUE\ archives for white dwarf companions of the nearest
Cepheids.

A complete analysis of our \IUE\ sample of late-type stars will be reported
in a subsequent paper.

\section{Observations}

Table 1 lists the parameters of the two F star targets.
Neither star has
a parallax measurement, and our source for the distances given in Table 1 will
be discussed in detail in the text.
As pointed out by \markcite{lsb93} Landsman \ea\ (1993),
knowledge of the distance of
the primary is essential for constraining the white dwarf parameters,
since low-dispersion \IUE\ spectra cannot be used to constrain
both \teff\ and \logg\ in the white dwarf.
The Str\"omgren photometry
used in the discussion of the distance determinations is
taken from \markcite{hauck90} Hauck and Mermilliod (1990).

The IUE images used in this study are listed in Table 2.
The standard IUESIPS
processing was used with the exception of the following three steps.
The fluxes were corrected for the long term degradation of the
sensitivity
of the SWP camera using a linear extrapolation of the tabulation of
\markcite{bohl88} Bohlin and Grillmair (1988).
The white-dwarf based absolute calibration was taken from Finley
(1993, personal communication).
The diffuse geocoronal and interplanetary \lya\ emission was removed
using the spatial information perpendicular to the dispersion in the IUE
line-by-line image (\cite{land93}).

The large-aperture spectra of each target were added together, after weighting
by the exposure time.
The gross flux of the small-aperture image SWP 52391 is negative
at wavelengths less than 1400 \AA, indicating that the
pedestal level for this image is below that of the Intensity Transfer
Function (ITF) used to linearize IUE fluxes.   The photometry of this image
should be considered uncertain due to the extrapolation of the ITF to negative
flux values (De La Pe\~{n}a 1994, personal communication), and thus it was
not used
for the co-added spectra.   The problem with negative gross fluxes also occured
to a lesser extent with the image SWP 52365, but this image was already given
low weight due to its short exposure time.

\section{Results}

\subsection{56 Per}

The co-added SWP spectrum of 56 Per is shown in Figure 1.   Also shown for
comparison is a spectrum of Procyon (F5 IV-V) scaled by a factor of 142, in
agreement with the difference of the V magnitudes.     This scaling of
Procyon provides an excellent match to the LWP (2000 - 3200 \AA)
spectrum of 56 Per.
The F star flux drops rapidly shortward of the Si II ionization edge at
1680 \AA, and the presence of a hot companion in 56 Per is indicated by
the persistent hot continuum toward shorter wavelengths.
The broad \lya\ absorption profile shows that the hot
companion is, in fact, a white dwarf.
Weak stellar \lya\ emission from the F star is detected in the core of the
white dwarf \lya\ profile.    The weak broad absorption feature at
1400 \AA\ indicates the presence of a quasi-molecular Lyman
alpha H-H satellite, commonly seen in warm (8000 - 17,000 K) DA white dwarfs
(e.g.\ \cite{berg95b}).

56 Per is a visual binary (ADS 3188) with a V $\sim 8.7$ variable star
located $4.3''$ from the primary.   The visible companion is too
luminous to be the white dwarf seen in the IUE spectra, and therefore 56 Per
must be (at least) a triple system.
Our large (10$''$ by 20$''$) aperture IUE observations do
not, by themselves, exclude the possibility that the white dwarf is
associated with the visible companion rather the F star primary.
This is because the orientation of the large aperture (P.A. = $143\arcdeg$)
was such that the visible companion was located predominantly in the direction
of the dispersion.
However, the ratio of the
flux at 1300 \AA\ (primarily due to the white dwarf) to the flux at 1700 \AA\
(primarily due to the F star) is approximately equal for both the small
($3''$ circle) and large aperture spectra.     Thus the white dwarf must be
located within about $1''$ from the F star.

Figure 2 shows the spectrum of 56 Per after subtraction of a spectrum of the
template star, Procyon.
Also shown is a spectrum (SWP 28185) of the white dwarf G148-7, scaled by a
factor of 1.45.   The G148-7 spectrum provides an excellent fit to the 56 Per
white dwarf spectrum, including the region of the \lya\ satellite at 1400 \AA.
{}From simultaneous fitting of the Balmer line profiles,
\markcite{berg95a} Bergeron et al.\ (1995a) derived \teff\ = 15480 K,
and \logg\ = 7.97 for G148-7.
However, the similarity of the ultraviolet spectra does {\em not} imply that
the white dwarfs have the same temperature and gravity (\cite{berg95b}).
Thus we follow \markcite{lsb93} Landsman et al.\ (1993), and tabulate
models for a grid of (\teff, \logg) values consistent with the IUE spectrum
of 56 Per (Table 3).
The models are computed as in \markcite{berg95b} Bergeron et al. (1995b),
with the ML2/$\alpha$ = 0.6
parameterization of the mixing-length theory, and the \lya\ quasi-molecular
satellite profiles of Allard \ea\ \markcite{all94}(1994).
The white dwarf mass is derived from the temperature and gravity
using the white dwarf cooling models of \markcite{wood95} Wood (1995)
for a pure carbon composition, with thick hydrogen and helium layers.
Note that the model with \teff = 15,460 K, \logg = 8.0 provides a very
close match to the parameters derived optically for G148-7 by
\markcite{berg95a}
Bergeron \ea\ (1995a), and illustrates the internal consistency of the
model parameters derived from IUE and optical spectroscopy.

A determination of the stellar distance is required to further constrain the
white dwarf parameters.
The Str\"omgren absolute magnitude calibration of
\markcite{niss88} Nissen (1988), as implemented
in the FORTRAN program of \markcite{nap93} Napiwotzki et al.\ (1993),
gives \mv\ = 3.37, and thus a
distance of 30.1 pc to 56 Per.
This absolute magnitude is in excellent
agreement with the value of \mv\ = 3.3, tabulated by \markcite{corb84}
Corbally and Garrison (1984)
for an F4 V star.   An error of one spectral type at F4 V corresponds to a
difference of 0.2 in \mv, or a difference of 2.8 pc in the distance estimate.
Interpolating a distance of $30.1 \pm 2.8$ pc in the grid in Table 3,
gives  \teff\ = $16,420 \pm 420$ K and \logg\ = $8.46 \pm 0.2$ for the white
dwarf, and an implied mass of $0.90 \pm 0.12$ \msun.

The mass distribution for single white dwarfs has a median value of
$\sim 0.6$ \msun, with a sparsely populated high-mass tail extending to
$\sim 1.0$ \msun\ (\cite{berg95b,brag95}).    According to current estimates
of the initial mass - final mass relation (IMFMR), this high mass tail has
its origin in progenitor stars with M $\gtrsim 2.5$ \msun, with the remnant
mass increasing smoothly with the progenitor mass above this threshold
(\cite{weid87,brag95}).     Most functional forms of the IMFMR
require a mass greater than  5 \msun\ for  the progenitor of
a 0.9 \msun\ white dwarf (c.f.\ Figure 23 in \markcite{wood92}
Wood 1992).
If the white dwarf progenitor mass were this large, it
would place a strong upper limit on the lifetime of the 56 Per system.
The pre-WD lifetime of a 5 \msun\ star is about 110 Myr (\cite{schal92}), and
the time required for a 0.9 \msun\ white dwarf to cool to \teff = 16,400 K
is 320 Myr (\cite{wood95}),  so that the total lifetime of the system must
be less than about 430 Myr.    Such a short lifetime seems implausible for
56 Per, since it is not identified with a young cluster or supercluster.
Unfortunately, since the F star in 56 Per is located near the main-sequence,
its Str\"omgren photometry is consistent with any evolutionary age up to
about 2 Gyr, according to the solar metallicity isochrones of \markcite{bert94}
Bertelli \ea\ (1994).

The SIMBAD database gives four radial velocity references for 56 Per, with
the average velocity falling between $-31.2$ and $-35.4$ \kms.
The radial
velocity is marked as variable in an early (1923) study, but no orbit has
been determined.    Since 56 Per appears to have a large mass ratio
($\sim 1.4$ \msun\ for the F4 V star, and $\sim 0.9$ \msun\ for the white
dwarf), the low velocity amplitude indicates either a low orbital inclination,
or, more likely, a long-period orbit.   Further radial velocity studies on 56
Per are clearly warranted.

\subsection{HR 3643}

Figure 3 shows the co-added IUE spectrum of HR 3643, along with the spectrum
of the template star $\upsilon$ Peg (F8 IV, \bv = 0.60).   The HR 3643
spectrum has been dereddened by E(\bv) = 0.04.    This reddening value, and
our use of $\upsilon$ Peg as a template star, are discussed further below.
The spectrum of HR 3643 shows a steeper rise of the UV continuum and a
narrower \lya\ absorption profile than does the 56 Per spectrum, indicating
a higher temperature white dwarf.   The very steep flux rise shortward
of \lya\ may indicate some
contribution by long-wavelength scattered light.    The spectrum shows
strong \oi\ $\lambda$ 1305 and possible \civ\ $\lambda$ 1550 emission,
but no other chromospheric line (including \lya) is clearly detected.

No evidence for radial velocity variations have been found
for HR 3643, which, in fact, has occasionally been adopted as a radial
velocity standard (e.g. \cite{layd94}).
\markcite{houk75} Houk and Cowley (1975) give a
spectral type of F7 II for HR 3643 and quote an unpublished type of F8 II
from Garrison and Hagen.     SIMBAD gives a best spectral type of F6 II-III
from \markcite{deVauc57} deVaucouleurs (1957), while the Yale Bright Star
Catalog (\cite{hoff91})
gives a spectral type of F9 II.    The bright giant classification
of HR 3643  thus seems secure, and provides a crude estimate of the absolute
magnitude.  \markcite{corb84} Corbally and Garrison (1984)
tabulate \mv $= -2.0$ for a F7 II star.
To estimate the possible range, we assume that HR 3643 is brighter than
a F7 III star (\mv $= 0.6$) and fainter than the F7~Ib-II low-amplitude
Cepheid, $\alpha$ UMi, which has \mv $= -2.94$ (\cite{fern95}).    These
absolute
magnitudes are used to derive the crude HR 3643 distance estimate of
60 -- 300 pc given in Table 1.

We are unable to determine a more refined distance to HR 3643 due to the
relative rarity of F bright giants at known distances.
For example, the  Str\"omgren F star absolute magnitude calibrations of
\markcite{craw75} Crawford (1975) or
\markcite{niss88} Nissen (1988) specifically exclude luminosity class II stars
with the Str\"omgren parameters of HR 3643
($\beta = 2.625$ and $\delta c_{1}$ = 0.26).
On the other hand, the Str\"omgren F supergiant standards used by
\markcite{gray91} Gray (1991)
or \markcite{arell90} Arellano Ferro and Parrao (1990) all appear to be more
luminous than HR 3643.
\markcite{egg89} Eggen (1989)
classified HR 3643 as an old disk giant and on that basis
 derived an absolute magnitude \mv\ = +1.4.
The kinematics of HR 3643 do place it just outside the region of young disk
stars (see Figure 3 in Eggen).    However, several pieces of evidence suggest
that HR 3643 is instead an intermediate mass star ($\sim 2 - 5$ \msun) with
a higher intrinsic luminosity.
First, the rotational velocity (\vsini  = 53 \kms) of HR 3643 is much larger
than typically found for old disk giants.    Second, the linewidths of the
Ca H and K profiles of HR 3643 are extremely broad, indicating
a high stellar luminosity from the Wilson-Bappu relation.    From Figure 2
in \markcite{drav81} Dravins (1981), after a crude correction for the
rotational velocity, we
estimate a Ca K linewidth of $\log W = 2.3$ \kms, which
is typical of a type Ib supergiant (\cite{wilson76}).
Additional evidence for a large intrinsic luminosity comes from the IUE
emission
line spectrum.   Figure 3 shows that \oi\ $\lambda$ 1305 emission is present,
but that \cii\ $\lambda$ 1335 and \civ\ $\lambda$ 1548 are either absent
or marginally detected.    The presence of strong \oi\ emission with weak
or absent \cii\ and \civ\ is characteristic of stars on the luminous
side of the chromosphere - transition region dividing line (\cite{linsky79}).
We  measure an \oi\ flux in HR 3643 of $3.9 \times 10^{-13}$ erg cm$^{-2}$
s$^{-1}$, and derive a normalized emission line flux of
f(\oi)/\lbol = $9 \times 10^{-7}$, which is typical of the values seen in
Cepheid stars (\cite{schmidt82}).
(HR 3643 is almost certainly not a Cepheid, since no radial
velocity variations have been found, and \markcite{fern76} Fernie (1976)
reports no evidence for photometric variability.)

A large distance to HR 3643 is also indicated by the evidence for
non-negligible reddening.
The reddening derived from
the the Str\"omgren F star calibration of
\cite{craw75}
is E(b--y) = 0.029, while the supergiant F star calibration of Gray
gives E(b--y) = 0.043.    \markcite{bers95} Bersier (1995) derived
E(\bv) = 0.071 toward HR 3643 on the basis of Geneva photometry.
We adopt E(\bv) $\sim$ E(b--y)/0.73 = 0.04 from the Crawford F star
calibration.
The implied hydrogen column density is then N(\hi) $ \sim 2 \times 10^{20}$
cm$^{-2}$, assuming a
dust to gas ratio of N(\hi)/E(\bv) = $5.2 \times 10^{21}$ cm$^{-2}$ mag$^{-1}$
(\cite{shull85}).   This large hydrogen column density would explain
the non-detection of the HR 3643 white dwarf in the
ROSAT WFC and EUVE surveys (\cite{pye95,bow95b}), and is also
consistent with our failure to detect any stellar \lya\ emission from the F
star in the core of the white dwarf \lya\ absorption profile
(\cite{land93}).

Figure 4 shows that the LWP spectrum of HR 3643 is well-matched by
a spectrum (LWP 14389) of $\upsilon$ Peg (F8 IV, V = 4.40, \bv = 0.60),
multiplied by the ratio (1.1) of the dereddened V magnitudes.
Among the late F stars with well-exposed SWP and LWP spectra in the
IUE archives, $\upsilon$ Peg provides the best fit
\footnote{Note that the IUE spectrum of HD 160365, adopted as the F6 III
standard  star in the IUE Spectral Atlas (\cite{wu83}) cannot be used as a F
star  template, because HD 160365 itself has a hot white dwarf companion
(\cite{bohm92}).}
to the LWP spectrum of HR 3643.
However, $\upsilon$ Peg is not an ideal template star because it is
considerably less luminous than HR 3643, and exhibits intense \cii\ and \civ\
chromospheric emission (\cite{simon89}).   In fact, Figure 3 shows that
the SWP spectrum  of $\upsilon$ Peg overcorrects for the contribution of the
F star near 1850 \AA.   Therefore, to create the HR 3643 white dwarf spectrum
shown in Figure 5, we scaled the $\upsilon$ Peg spectrum by a factor (0.7)
chosen to yield a smooth continuum following the subtraction.    We then fit
white dwarf models to the spectrum shortward of 1700 \AA.   However, we would
have derived very similar white dwarf parameters, if we had fit only to
wavelengths $<$ 1500 \AA\, where the contribution of the F star is negligible.

Table 3 displays a grid of white dwarf models consistent with the HR 3643
white dwarf spectrum, and Figure 5 shows a model spectrum for the case of
\logg = 8.0.
The distance estimate for HR 3643 given in Table 1
is too uncertain to provide any further constraint on the white dwarf
parameters.    Instead, the absolute magnitude for the F
star can be constrained, by assuming a reasonable range of surface gravity
for the white dwarf.   The observed surface gravities for
isolated white dwarfs range between $7.3 <$ \logg $< 9.0$ ({\cite{berg92}).
{}From Table 2, this surface gravity constraint implies a
a distance to HR 3643 of 56 $<$ d $<$ 150 pc, and an absolute magnitude of the
F star of $-1.4 <$ \mv $< 0.6$.
Should the distance to HR 3643 turns out to be greater than $\sim $ 150 pc,
then the possibility must be considered that the white dwarf is not physically
associated with HR 3643.  However, the spatial separation of the two stars
must be less than about $2''$, since the line-by-line IUE spectra
do not show a  spatial shift between the regions of the spectrum dominated
by the white dwarf and the F star.   Given this close proximity, and the
broad overlap in the estimated distances of the F star and the white dwarf,
we consider a physical association to be highly probable.

HR 3643 may be the most massive star known to have a white
dwarf companion, and thus is an important object for the study
of the massive progenitors of white dwarfs.
However, further analysis of this system will require a better
distance determination, for example, from the results of the
{\em Hipparcos} satellite.     Another important observation will be to
obtain {\em high}--dispersion \lya\ observations, which,
unlike the case with the low-dispersion \IUE\ \lya\ profile, can be used to
constrain both the gravity and temperature of the white dwarf
(\cite{lsb95}).

\section{White Dwarf Companions to Cepheids}

The mass of the white dwarf progenitor in a non-interacting binary must have
been larger than that of the remaining non-degenerate star.    Thus, in
principle, binary systems can provide constraints on
the upper mass, M$_u$, of a star that can leave a white dwarf remnant,
and on the initial mass - final mass relation (IMFMR) that accounts for the
total mass loss during stellar evolution.
The detection of a close white dwarf companion to a main-sequence B
star is not possible using ultraviolet spectroscopy
\footnote{In principle, hot white dwarf companions to B stars could be
detected with EUV spectroscopy, but no such systems are yet known.   Also,
white dwarf companions to B stars may have been detected in interacting
systems such Be stars (e.g.\ \cite{haberl95}).}.
Thus, our empirical knowledge of M$_u$ and the IMFMR is based mainly on
observations of young open clusters (\cite{weid90}), where the white dwarf
progenitors must have been more massive than the turn-off mass of the cluster.
In particular, the detection of massive white dwarfs in the young cluster
NGC 2516 shows that stars with masses up to 8($^{+3}_{-2}$) \msun\ can leave
white dwarf remnants (\cite{reim82}).

After a main-sequence B star evolves into a cool supergiant, the detection
of a hot white dwarf companion becomes feasible in the ultraviolet.    In the
case of HR 3643, the main-sequence progenitor of the F bright giant
(with \teff\ = 6000 K and \mv\ = $-0.4 \pm 1$) was probably a late B star of
mass 2--4 \msun, according to the evolutionary tracks of
\markcite{schal92} Schaller \ea\ (1992).    However, the case of HR 3643 also
illustrates the difficulty in obtaining accurate values of the
distance, mass, and age of non-variable bright giants or supergiants.
For example, Gray (1991) found that the scatter in absolute magnitude
for non-variable F supergiants was too large to allow a useful calibration with
Str\"omgren photometry.   Some of this scatter is due to the fact that the
helium-burning evolutionary tracks of intermediate-mass stars can
overlap, so that a single photometric box or MK spectral type does not
correspond to a unique mass.

In contrast, the discovery of a Cepheid with a white dwarf companion would
provide
particularly useful constraints on M$_u$ and the IMFMR.
The minimum mass for a Cepheid is 3--4 \msun, so the white dwarf
progenitor must be at least this massive.    More importantly, the
the Cepheid pulsations can be used to estimate the distance and age
of the system.     Comparison of the white dwarf cooling age with the
evolutionary age of the Cepheid would then provide a relatively precise
estimate of the mass of the white dwarf progenitor (c.f.\ \cite{evans94}).
The time for a white dwarf to cool to
15,000 K is 540 Myr for a 1.0 \msun\ remnant, and 210 Myr for 0.6 \msun\
remnant (\cite{wood95}).   On the other hand, the hydrogen burning lifetime of
a 4 \msun\ star (the progenitor of a low-mass Cepheid) is about 165 Myr
(\cite{schal92}).
Thus, if a Cepheid originally had a more massive companion, then the white
dwarf remnant of the companion, if it exists, should still be detectable
by deep ultraviolet spectroscopy.

What percentage of Cepheids should have white dwarf companions?
The original mass ratio, $q = M_2/M_1$ for such a system must be
greater than about 0.5, since the primary, M$_1$ must be less massive than
M$_u$ ($\sim 8$ \msun),
and the secondary, M$_2$, must be more massive than the minimum Cepheid mass
($\sim$ 4 \msun).    The system must also have a long-period orbit
($>$ 1 year) so
that each star can evolve to a supergiant without interacting with its
companion.  \markcite{abt90} Abt \ea\ (1990)
found that B stars in long-period orbits did
{\em not} favor  mass ratios near unity, and instead were distributed as a
power law toward lower masses.     However, it is plausible
that at least some Cepheid-white dwarf binaries exist, since,
out of 20 Cepheids with well-determined orbits,
Evans \markcite{evans95}(1995)
found four with a secondary masses between 4 and 5.5 \msun.    Such systems
could eventually evolve into a Cepheid - white dwarf binary, if the current
secondary eventually evolves into a Cepheid, and the current Cepheid leaves
a white dwarf remnant.

No white dwarf companion to a Cepheid has yet been found, despite the fact that
numerous Cepheids have been observed with \IUE.
For example, \markcite{evans92}Evans (1992) has conducted a
magnitude-limited survey to 8th magnitude of 76 Cepheids with the
long-wavelength camera on IUE, while \markcite{evans95}Evans (1995)
has obtained deep
short-wavelength images of 20 Cepheids with well-determined orbits.
However, because
her surveys were designed to detect main-sequence companions, they
concentrated on longer wavelengths (and shorter exposure times) than is
optimal for a white dwarf
search.   Therefore, we have examined the archival IUE data set for the nearest
Cepheids in order to assess the current limits on white dwarf companions.

Table 4 lists the 15 Cepheids within 500 pc, as tabulated in the
catalog of \markcite{fern95} Fernie \ea\ (1995).
(EW Sct has been omitted from the list due
to its very large reddening, E(\bv) = 1.1.)  For each star, we extracted
the image with the longest SWP exposure.
We used NEWSIPS processed data (\cite{nichols94}) whenever it was available.
Two of the Cepheids in Table 4 (BG Cru and DT Cyg) have not been observed
with the SWP camera.   Another three Cepheids (W Sgr, SU Cas, and $\eta$ Aql)
have hot ($>$ 9,000~K) main-sequence companions (Evans 1992) which swamp any
potential signal from a white dwarf in the IUE wavelength range.

We examined each image for evidence of a broad \lya\ absorption, but found no
white dwarf candidates.   Several of the deepest images show a very steep
rising flux shortward of \lya, which is a signature of long-wavelength
scattered light.     We removed this scattered light by subtracting a constant
level of IUE flux units so that the signal below \lya\ (where the IUE
sensitivity rapidly decreases) goes to zero.
We compute an upper limit to the flux in a 50~\AA\ bandpass centered
at 1345~\AA, by adding the residual mean flux in this bandpass
to the 1 sigma flux variations within the bandpass.
This wavelength is chosen as  optimal for a white dwarf search, because the
contribution  from the Cepheid at 1345~\AA\ is expected to be small, and the
contribution of the white dwarf will not be strongly affected by a possible
broad \lya\ absorption.

We then used white dwarf atmosphere models to determine the temperatures of a
0.6~\msun\ and a 1.0~\msun white dwarf consistent with the flux upper
limits, using the distances and reddenings tabulated in Table 4.
Although the white dwarf mass function peaks near 0.6~\msun\
(\cite{brag95}), the 1.0~\msun\ model may be more appropriate
for the remnant of a massive ($>3$ \msun) progenitor.
The reddening of the model UV flux was determined from E(\bv) using the
parameterization of \markcite{card89} Cardelli \ea\ (1989).
The white dwarf radii, needed to determine the flux scaling at the Cepheid
distance, were computed from the mass and temperature using the
cooling models of \markcite{wood95} Wood (1995).

According to Table 4, the upper limit on the 1345 \AA\ flux  from
$\alpha$ UMi (Polaris) implies that a 0.6 \msun\ white dwarf companion must be
be cooler than 14,370 K,  and a 1.0 \msun\ companion must be cooler than
16,840 K.    The corresponding white dwarf cooling ages are, respectively,
240 Myr and 400 Myr.     As noted above,
these cooling ages are longer than the lifetime of the Cepheid, so that
a white dwarf companion of $\alpha$ UMi is ruled out by the IUE observations.
Since $\alpha$ UMi is a well-known astrometric and spectroscopic binary,
its companion is thus likely to be an F star \markcite{evans88,evans95}
(Evans 1988, 1995).     A white dwarf companion can probably also be ruled
out for $\beta$ Dor, for which the upper limit in Table 4 of \teff\
$<$ 19,080 K for a 0.6 \msun\ white dwarf  implies a  cooling age greater
than 90 Myr.   The absolute magnitude of $\beta$ Dor is $-4.08$
(\cite{fern95}),
corresponding to an evolutionary mass of about 7 \msun\
and a lifetime of less than 60 Myr, according to the
evolutionary tracks of \markcite{beck77} Becker \ea\ (1977).
The use of evolutionary tracks with larger
convective core overshoot would give a smaller mass and somewhat longer
lifetime for $\beta$ Dor \markcite{evans95} (c.f.\ Evans 1995), but are
still unable to accomodate a white dwarf cooling age greater than
90 Myr.

For the remaining Cepheids listed in Table 4,
the existing IUE data allow for the possibility of a white dwarf
cooler than the upper limit on \teff,  but still young enough to be consistent
with the Cepheid evolutionary age.
Deeper  ultraviolet observations of these Cepheids are needed to reveal or rule
out the presence of a white dwarf companion.

\section{Summary}

We have detected white dwarf companions to the
F4 V star, 56 Per, and the F7 II star, HR 3643, using IUE spectroscopy.
We derive \teff\ = $16,420 \pm 420$ K and logg = $8.46 \pm 0.2$ for
the white dwarf companion to 56 Per, using the photometric distance of
$30.1 \pm 2.8$ pc.
The implied white dwarf mass is $0.90 \pm 0.12$ \msun.
56 Per is a known wide ($4.3''$) binary, but a small-aperture
\IUE\ spectrum is used to show that the white dwarf is a close companion of the
F star, and thus that 56 Per is a triple system.

The parameters of the white dwarf in HR 3643 are not well constrained, mainly
due to a large uncertainty in the distance of the primary.
By assuming a reasonable range of gravity, ($7.3 <$ \logg $< 9.0$),
for the white dwarf, we derive $-1.4 <$ \mv $< 0.6$ for the F7~II star,
and $28,970 <$ \teff $< 35,990$ K for the white dwarf.

Neither star has a parallax measurement, and the results from {\em Hipparcos}
will provide important physical constraints.   For 56 Per, an accurate
parallax is needed to confirm the high white dwarf mass implied by the
photometric distance.    In the case of HR 3643, a parallax is needed because
photometric and spectroscopic distances  lack sufficient precision to provide
a useful constraint on the white dwarf parameters.

Prompted by our detection of a white dwarf companion to a F7 II star,
we study the feasibility for the ultraviolet detection of white dwarf
companions to Cepheids.     The detection of a
Cepheid -- white dwarf binary would provide important insights into the
most massive progenitors of white dwarfs.
The Cepheid distance can be used along with ultraviolet spectroscopy to
determine the temperature and gravity of the white dwarf.
The progenitor of the white
dwarf must have been more massive than the existing Cepheid ($>$ 3 \msun),
and the evolutionary
age of the Cepheid would allow a fairly precise  mass estimate for the
progenitor.
{}From consideration of the evolutionary times,  we show the white dwarf
remnant of a Cepheid companion, if it exists,
should still be hot ($>$ 15,000 K), and detectable by ultraviolet
spectroscopy.  Only for the cases of $\alpha$ UMi and $\beta$ Dor,
are  existing \IUE\ spectra sufficiently deep to rule out the presence of a
white dwarf companion.

\acknowledgments

We thank Yoji Kondo and the staff of the \IUE\ Observatory for their
assistance in the acquisition of these data.    We also thank Richard
Gray and Nancy Evans for their comments concerning the absolute magnitude
calibration of F bright giants.
This research has made use of the SIMBAD database, operated at
CDS, Strasbourg, France and of the HEASARC
Online Service, provided by the NASA-Goddard Space Flight Center.
This research was supported in part by NASA IUE grant S-14636-F to Hughes
STX Corporation, by the NSERC Canada, and by the
Fund FCAR (Qu\'ebec).

\clearpage

\tightenlines

\begin{deluxetable}{rrrrrrr}
\tablecaption{Stellar Parameters}
\tablehead{
 \colhead{HR}  & \colhead{Name} & \colhead{HD} &  \colhead{Spec}  &
\colhead{V} & \colhead{\bv}   &   \colhead{d (pc)} }
\startdata
 1379 & 56 Per & 27786 & F4 V  & 5.76 & 0.40 & $30.1 \pm 2.8$  \\
 3643 &        & 78791 & F7 II & 4.48 & 0.61 & $60 - 300$   \\
\enddata

\end{deluxetable}

\clearpage

\begin{deluxetable}{llllrll}
\tablecaption{\IUE\ Observing Log}
\tablehead{
    \colhead{Star} & \colhead{Year} & \colhead{Day} & \colhead{Image}
   & \colhead{Exp (s)} & \colhead{Aperture} & \colhead{Comment} }
\startdata
       &     &      &           &    &   & \\
56 Per & 94  & 286  & LWP 29387 & 90 &  L & \\
       & 94  & 286  & LWP 29388 & 30 &  L & \\
       & 94  & 283  & SWP 52365 & 900 & L & Negative fluxes \\
       & 94  & 286  & SWP 52391 & 2700 & S & Negative fluxes \\
       & 94  & 286  & SWP 52392 & 2400 & L & \\
       & 94  & 286  & SWP 52393 & 1800 & L & \\
\cutinhead{}
HR 3643  & 94  & 318 & LWP 29509 & 3360 & L & Multiple \\
         & 94  & 309  & SWP 52733 &  900  & L &   \\
         & 94  & 318  & SWP 52797 &  3600 & L &  \\
\enddata
\end{deluxetable}

\clearpage

\begin{deluxetable}{lllllll}
\tablecaption{White Dwarf Model Fits}
\tablehead{
\colhead{} &  \colhead{\logg} & \colhead{\teff\ (K)} &
\colhead{R$^2$/D$^2$} & \colhead{M/\msun} & \colhead{\mv} & \colhead{d (pc)}}
\startdata
56 Per & 7.50  & 14520  & $9.08 \times 10^{-23}$ & 0.435  & 10.61  & 43  \nl
         &8.00  & 15460 & $6.50 \times 10^{-23}$ &0.615  & 11.20 &  36   \nl
         &8.50  & 16515  & $4.63 \times 10^{-23}$ &0.931  & 11.88 &  30  \nl
         &9.00  & 17620  & $3.56 \times 10^{-23}$ &1.198  & 12.72 &  22  \nl
\cutinhead{}
HR 3643 &  7.00  & 26900 &$1.32 \times 10^{-23}$    & 0.262  & 8.55  &  176 \nl
         & 7.50  & 28970  & $9.81 \times 10^{-24}$  & 0.431  & 9.23  &  139 \nl
        & 8.00  & 31130 & $7.62 \times 10^{-24}$  & 0.652 & 9.87 &  109  \nl
        & 8.50  & 33250   & $6.23 \times 10^{-24}$  & 0.952  & 10.58 &  82 \nl
       & 9.00  & 35990   & $5.20 \times 10^{-24}$  & 1.188  &11.46 & 56 \nl
\enddata
\end{deluxetable}

\clearpage
\begin{deluxetable}{rllccrrrrrrc}
\scriptsize
\tablecolumns{12}
\tablecaption{Limits on White Dwarf Companions to Cepheids}
\tablehead{
\colhead{HD} & \colhead{Name} & \colhead{$<$V$>$} & \colhead{$<$B$>$--$<$V$>$}
 & \colhead{E(\bv)} & \colhead{d}  & \colhead{SWP} & \colhead{Time} &
 \colhead{Flux\tablenotemark{a}} & \multicolumn{2}{c}{\teff (K)} &
 \colhead{Note} \\
\colhead{}   & \colhead{}  & \colhead{} & \colhead{} & \colhead{} &
  \colhead{(pc)}     & \colhead{}    & \colhead{(min)} &
  & \colhead{0.6 \msun} & \colhead{1.0 \msun} & \colhead{} }
\startdata
8890  & $\alpha$ UMi & 1.98 & 0.60 & 0.00 & 97 & 28557 & 65 & $<$ 1.29
      & $<$ 14370 & $<$ 16840 & \nl
17463 & SU Cas       & 5.97 & 0.70 & 0.29 & 265 & 16480 & 40  &
\nodata & \nodata & \nodata &  1 \\
29260 & SZ Tau       & 6.53 & 0.84 & 0.29 & 450 & 24985 & 120 & $<$ 0.79
      & $<$ 64400 & \nodata & 3 \nl
37350 & $\beta$ Dor  & 3.73 & 0.81 & 0.04 & 340 & 28452 & 400 & $<$ 0.36
      & $<$ 19080 & $<$ 23550 & \nl
45412 & RT Aur       & 5.45 & 0.59 & 0.05 & 430 & 7188  & 70 & $<$ 1.14
      & $<$ 28850 & $<$ 43400 & \nl
52973 & $\zeta$ Gem  & 3.92 & 0.80 & 0.02 & 390 & 17939 & 180 & $<$ 0.43
      & $<$ 20310 & $<$ 25320 & \nl
68808 & AH Vel       & 5.69 & 0.58 & 0.07 & 500 & 21379 & 120 & $<$ 1.03
      & $<$ 27800 & $<$ 40140 & \nl
108968& BG Cru       & 5.49 & 0.61 & 0.05 & 410 & \nodata & \nodata  &
\nodata & \nodata & \nodata & 2 \nl
161592 & X Sgr       & 4.55 & 0.74 & 0.20 & 330 & 6216  & 60 & $<$ 1.38
      & $<$ 41000  & $<$ 84950 & \nl
164975 & W Sgr       & 4.67 & 0.75 & 0.11 & 410 & 15303 & 29 &
\nodata & \nodata & \nodata & 1 \nl
176155& FF Aql       & 5.37 & 0.76 & 0.22 & 360 & 10085 & 160 & $<$ 0.73
      & $<$ 35850  & $<$ 67860  & \nl
180583 & V473 Lyr    & 6.18 & 0.63 & 0.03 & 365 &  8317 & 16 & $<$ 1.83
      & $<$ 28660  & $<$ 42780 & \nl
187929& $\eta$ Aql   & 3.90 & 0.79 & 0.15 & 270 &  5701 & 48 &
\nodata & \nodata & \nodata & 1 \nl
201078& DT Cyg       & 5.77 & 0.54 & 0.04 & 400 & \nodata & \nodata & \nodata &
\nodata & \nodata & 2 \nl
213306& $\delta$ Cep & 3.95 & 0.66 & 0.09 & 250 & 28556 & 300 & $<$ 1.10
      & $<$ 23650  & $<$ 30480 & \nl
\enddata

\tablenotetext{a}{$\times 10^{-14}$ erg cm$^{-2}$ s$^{-1}$ \AA$^{-1}$ }
\tablecomments{(1) Hot main-sequence companion present,
(2) No SWP observations exist,
(3) No \teff\ constraint possible for 1 \msun\ model }

\end{deluxetable}

\clearpage

\begin{figure}
\plotone{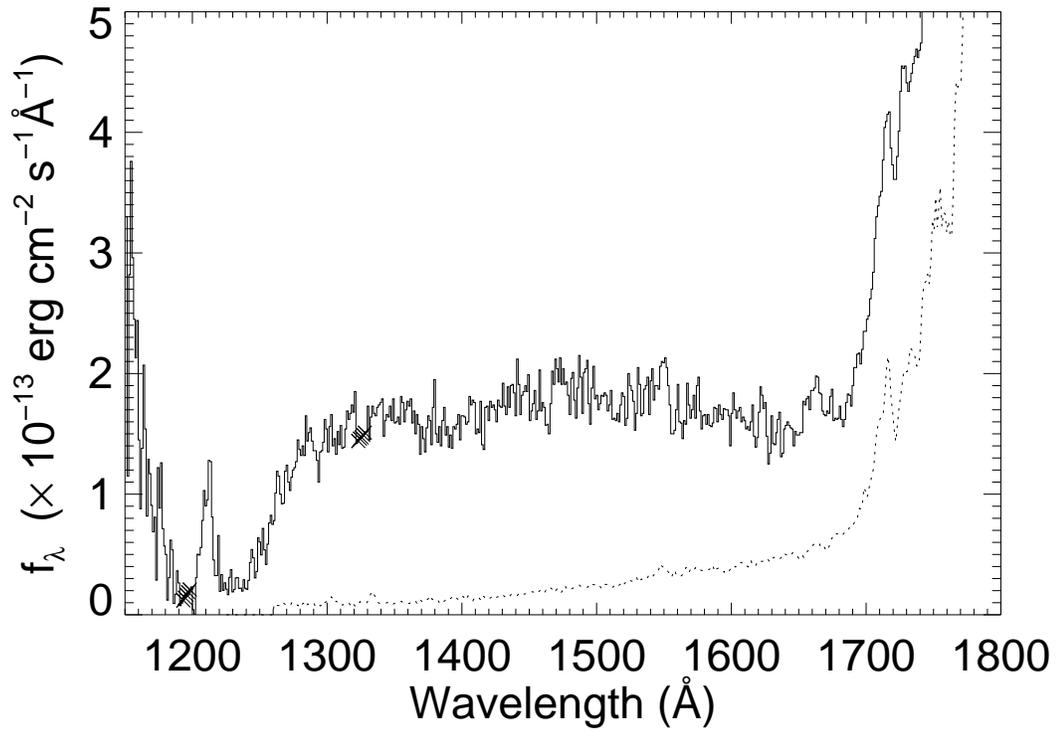}
\caption{The co-added IUE spectrum of 56 Per (solid line).   Geocoronal \lya\
emission has been removed as described in the text, and the observed \lya\
emission is chromospheric.   The dotted line
shows our adopted template spectrum of Procyon, scaled down by a factor of
142.}
\end{figure}

\begin{figure}
\plotone{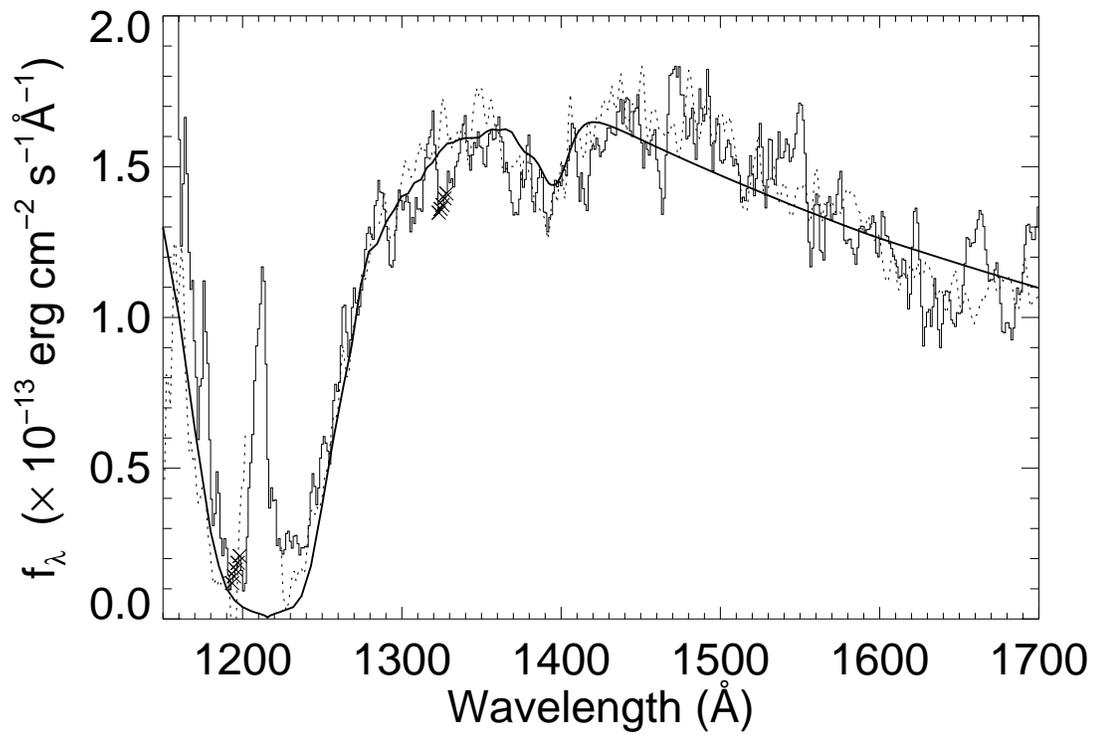}
\caption{The co-added IUE spectrum of 56 Per after subtraction of the template
star (solid line).    The thick solid line shows our best-fit model assuming
\logg\ = 8.5.   The dotted line shows the spectrum of the white dwarf
 G148-7 (SWP 28185) scaled by a factor of 1.45.}
\end{figure}

\begin{figure}
\plotone{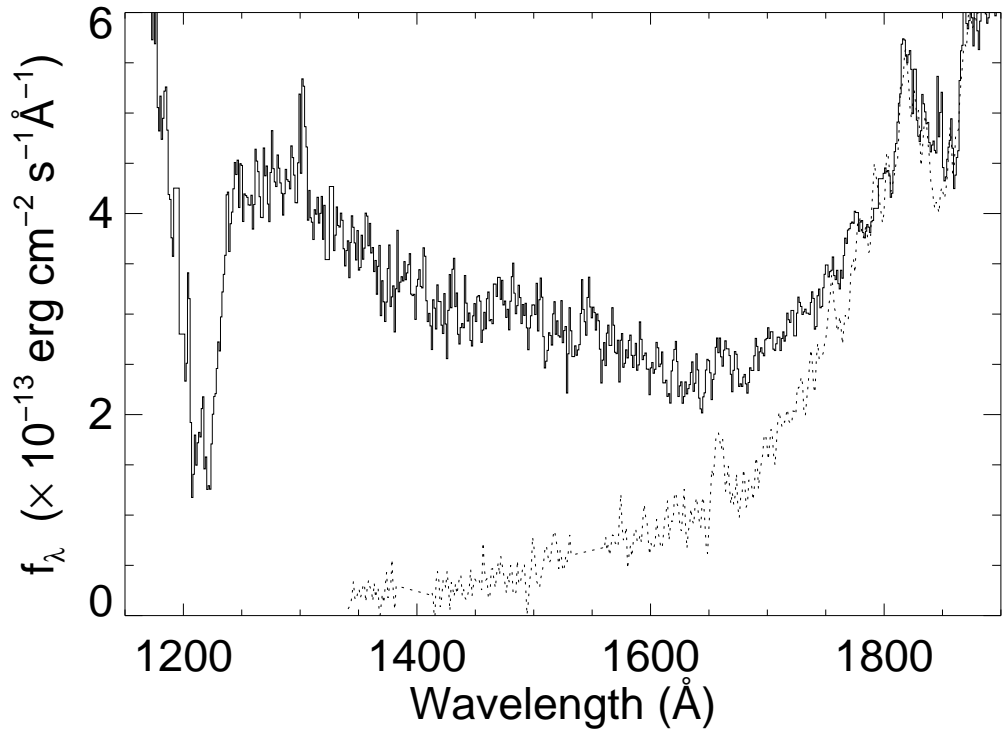}
\caption{The co-added IUE spectra of HR 3643 (solid line), after dereddening
with E(\bv) = 0.04.     The geocoronal
\lya\ emission has been removed as described in the text.   The dotted line
shows our template spectrum of Ups Peg (SWP 40364), multplied by the ratio
(1.1) of the dereddened V magnitudes.}
\end{figure}

\begin{figure}
\plotone{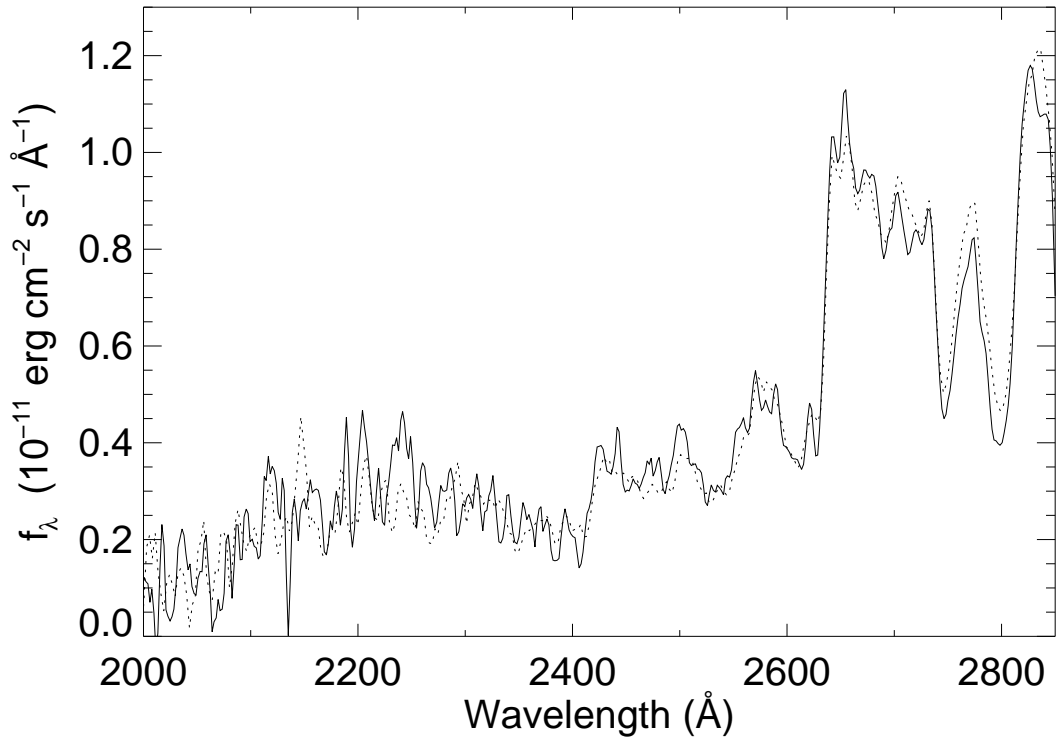}
\caption{The long-wavelength spectrum (LWP 29509) of HR 3643, dereddened with
E(\bv) = 0.04.  A model white dwarf spectrum (derived from fitting
the short wavelengths of the SWP spectrum) has been subtracted.
The dotted line shows the spectrum of $\upsilon$ Peg after scaling by the ratio
(1.1) of the dereddened V magnitudes.}
\end{figure}

\begin{figure}
\plotone{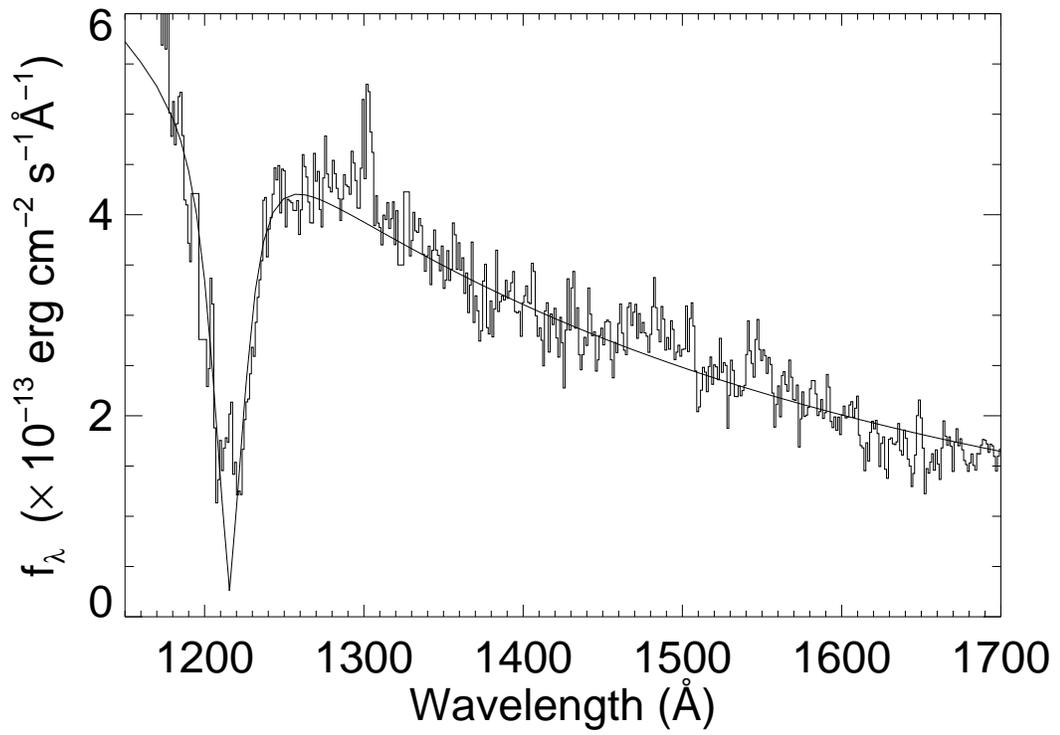}
\caption{The spectrum of HR 3643 after subtraction of 0.7 times the spectrum
of the template star, $\upsilon$ Peg.   The thick solid line shows a
best-fit white dwarf model, assuming \logg\ = 8.0}
\end{figure}

\end{document}